# Robustness of Regular Graphs Based on Natural Connectivity

Jun Wu[1,2*], Mauricio Barahona[2,3], Yuejin Tan[1], Hongzhong Deng[1]

*1 Department of Management, College of Information Systems and Management, National University of Defense Technology, Changsha 410073, P. R. China.*

*2 Institute for Mathematical Sciences, Imperial College London, London SW7 2PG, United Kingdom*

*3 Department of Bioengineering, Imperial College London, London SW7 2AZ, United Kingdom*

**Abstract**: It has been recently proposed that the natural connectivity can be used to characterize efficiently the robustness of complex networks. The natural connectivity quantifies the redundancy of alternative routes in the network by evaluating the weighted number of closed walks of all lengths and can be seen as an average eigenvalue obtained from the graph spectrum. In this paper, we explore both analytically and numerically the natural connectivity of regular ring lattices and regular random graphs obtained through degree-preserving random rewirings from regular ring lattices. We reformulate the natural connectivity of regular ring lattices in terms of generalized Bessel functions and show that the natural connectivity of regular ring lattices is independent of network size and increases with $K$ monotonically. We also show that random regular graphs have lower natural connectivity, and are thus less robust, than regular ring lattices.

---

[*] Tel.: +86-731-4573593; Fax: +86-731-4573593; E-mail: wujunpla@hotmail.com

# 1. Introduction

Networks are everywhere. Networks with complex topology describe a wide range of systems in nature and society (Albert and Barabási 2002, Boccaletti *et al.* 2006, Newman 2003). The investigation of complex networks has recently been one of the most popular topics within the interdisciplinary area involving physics, mathematics, biology, social sciences, informatics, and other theoretical and applied sciences. Complex networks rely for their function and performance on their robustness, i.e., the ability to endure threats and survive accidental events. Because of its broad application, the robustness of complex networks has received growing attention (Albert *et al.* 2000, Wu *et al.* 2007b, Wu *et al.* 2007a, Cohen *et al.* 2001, Holme *et al.* 2002, Cohen *et al.* 2000).

Simple and effective measures of robustness are essential for improved design in different areas of systems science. A variety of measures, based on different heuristics, have been proposed to quantify the robustness of networks. For instance, the vertex (edge) connectivity of a graph is an important, and probably the earliest, measure of robustness of a network (Whitney 1932). However, the connectivity only partly reflects the ability of graphs to retain connectedness after vertex (or edge) deletion. Other improved measures include super connectivity (Bauer and Bolch 1990), conditional connectivity (Harary 1983), restricted connectivity (Esfahanian and Hakimi 1988), fault diameter (Krishnamoorthy and Krishnamurthy 1987), toughness (Chvatal 1973), scattering number (Jung 1978), tenacity (Cozzens *et al.* 1995), the expansion parameter (Alon 1986) and the isoperimetric number (Mohar 1989). In contrast to vertex (edge) connectivity, these new measures consider both the cost to damage a network and how badly the network is damaged. Unfortunately, from an algorithmic point of view, the problem of calculating these measures for general graphs is NP-complete. This implies that these measures are of no great practical use within the context of complex networks. Another remarkable measure to unfold the robustness of a network is the second smallest (first non-zero) eigenvalue of the Laplacian matrix, also known as the algebraic connectivity. Fiedler (Fiedler 1973) showed that the magnitude of the algebraic connectivity reflects how well connected the overall graph is; the larger the algebraic connectivity is, the more difficult it is to cut a graph into independent components. However, the algebraic connectivity is equal to zero for all disconnected networks. Therefore, it is too coarse a measure for complex networks.

An alternative formulation of robustness within the context of complex networks emerged from random graph theory (Bollobás 1985) and was stimulated by the work of Albert et al. (Albert *et al.* 2000). Instead of a strict extremal property, they proposed a statistical measure, i.e., the critical removal fraction of vertices (edges) for the disintegration of a network, to characterize the robustness of complex networks. The disintegration of networks can be observed from the decrease of network performance. The most common performance measurements include the diameter, the size of the largest component, the average path length, the efficiency (Latora and Marchiori 2001) and the number of reachable vertex pairs (Siganos *et al.* 2006). As the fraction of removed vertices (or edges) increases, the performance of the network will eventually collapse at a critical fraction. Although we can obtain the analytical critical removal fraction for some special networks analytically (Cohen *et al.* 2000, Callaway *et al.* 2000, Cohen *et al.* 2001, Wu *et al.* 2007b), this measure can only be calculated through simulations in general.

Recently, we showed that the concept of natural connectivity can be used to characterize the robustness of complex networks (Wu *et al.* 2008). The natural connectivity is based on the Estrada index of a graph, which has been proposed to characterize molecular structure (Estrada 2000), bipartivity (Estrada and Rodríguez-Velázquez 2005a), subgraph centrality (Estrada and Rodríguez-Velázquez 2005b) and expansibility (Estrada 2006a, Estrada 2006b). The natural connectivity has an intuitive physical meaning and a simple mathematical formulation. Physically, it characterizes the redundancy of alternative paths by quantifying the weighted number of closed walks of all lengths leading to a measure that works both in connected and disconnected graphs. Mathematically, the natural connectivity is obtained from the graph spectrum as an average eigenvalue and it increases strictly monotonically with the addition of edges. Rich information about the topology and dynamical processes can be extracted from the spectral analysis of the networks. The natural connectivity sets up a bridge between graph spectra and the robustness of complex networks, allowing a precise quantitative analysis of the network robustness.

In this paper, we investigate the natural connectivity of regular graphs, i.e., graphs with constant degree, both of deterministic and of random nature. In particular, we study deterministic ring lattices (*K*-cycles or pristine worlds) and regular randomized graphs derived from them through a degree-preserving rewiring (Watts and Strogatz 1998, Barahona and Pecora 2002). Regular graphs are networks of wide interest and constitute examples of expander graphs, which are extremely robust to vertex or edge removal (Sarnak 2004, Hoory *et al.* 2006). The paper is structured as follows. In Section 2, we provide some basic graph definitions and state the concept of natural connectivity together with some fundamentals of multivariable generalized Bessel functions. These functions are used in Section 3 to study the natural connectivity of regular ring lattices through numerical and analytical methods. In Section 4, we discuss the natural connectivity of randomized regular graphs obtained from regular ring lattices. Finally, conclusions are presented in Section 5.

## 2. Preliminaries

### 2.1 Graphs and Natural Connectivity

A complex network can be formalized in terms of a simple undirected graph $G(V,E)$, where $V$ is the set of vertices and $E \subseteq V \times V$ is the set of edges. Let $N = |V|$ and $M = |E|$ be the number of vertices and the number of edges, respectively. The connectivity of the graph $G$ can be represented by the adjacency matrix $A(G) = (a_{ij})_{N \times N}$, where $a_{ij} = a_{ji} = 1$ if vertex $v_i$ and $v_j$ are adjacent and $a_{ij} = a_{ji} = 0$ otherwise. It follows immediately that $A(G)$ is a real symmetric matrix with real eigenvalues $\lambda_1 \geq \lambda_2 \geq ... \geq \lambda_N$, which are usually also called the eigenvalues of the graph $G$ itself. The set $\{\lambda_1, \lambda_2, ... \lambda_N\}$ is called the spectrum of $G$.

A walk of length $k$ in a graph $G$ is an alternating sequence of vertices and edges $v_0 e_1 v_1 e_2 ... e_k v_k$, where $v_i \in V$ and $e_i = (v_{i-1}, v_i) \in E$. A walk is closed if $v_0 = v_k$. The number of closed walks is an important index for complex networks. Recently, we have proposed that the number of closed walks of all lengths quantify the redundancy of alternative paths in the graph and can therefore serve as a measure of network robustness (Wu *et al.* 2008). Considering that shorter closed walks have more influence on the redundancy than longer closed walks, we scale the contribution of closed walks by dividing them by the factorial of the length $k$. That is, we define a weighted sum of numbers of closed walks $S = \sum_{k=0}^{\infty} n_k / k!$, where $n_k$ is the number of closed walks of length $k$. This scaling ensures that the weighted sum does not diverge and it also means that $S$ can be obtained from the powers of the adjacency matrix:

$$n_k = \sum_{i=1}^{N} \lambda_i^k = trace(A^k) = \sum \lambda_i^k \tag{1}$$

Using Eq. (1), we can obtain

$$S = \sum_{k=0}^{\infty} \frac{n_k}{k!} = \sum_{k=0}^{\infty} \sum_{i=1}^{N} \frac{\lambda_i^k}{k!} = \sum_{i=1}^{N} \sum_{k=0}^{\infty} \frac{\lambda_i^k}{k!} = \sum_{i=1}^{N} e^{\lambda_i} \tag{2}$$

Hence the proposed weighted sum of closed walks of all lengths can be derived from the graph spectrum. We remark that Eq. (2) corresponds to the Estrada Index of the graph (Estrada 2000), which has been used in several contexts in graph theory, including subgraph centrality (Estrada and Rodríguez-Velázquez 2005b) and bipartivity (Estrada and Rodríguez-Velázquez 2005a). The natural connectivity can be defined as an average eigenvalue of the graph as follows.

**Definition 1** (Wu *et al.* 2008) *Let $A(G)$ be the adjacency matrix of $G$ and let $\lambda_1 \geq \lambda_2 \geq ... \geq \lambda_N$ be the eigenvalues of $A(G)$, the natural connectivity or natural eigenvalue of $G$ is defined by*

$$\bar{\lambda} = \ln\left(\sum_{i=1}^{N} e^{\lambda_i} / N\right) \tag{3}$$

It is evident from Eq. (3) that $\lambda_1 \geq \bar{\lambda} \geq \lambda_N$. Moreover, the natural connectivity changes strictly monotonically with the addition or deletion of edges (Estrada and Rodríguez-Velázquez 2005a, Wu *et al.* 2008). Previous work that the natural connectivity provides a sensitive and reliable measure of the robustness of the graph (Estrada 2006b, Wu *et al.* 2008).

**2.2 Generalized Bessel Functions**

The Bessel functions of the first kind $J_\alpha(x)$ are defined as the solutions to the Bessel differential equation

$$x^2 \frac{d^2 y}{dx^2} + x\frac{dy}{dx} + (x^2 - \alpha^2)y = 0 \tag{4}$$

which are nonsingular at the origin (Abramowitz and Stegun 1972). Another definition of the Bessel function for integer values of $n$ is possible using an integral representation (Abramowitz and Stegun 1972)

$$J_n(x) = \frac{1}{\pi}\int_0^\pi \cos(n\tau - x\sin\tau)d\tau \tag{5}$$

The generalized Bessel functions of the first kind for integer values of $n$ are defined by (Dattoli *et*

*al.* 1991):

$$J_n(x_1, x_2, ...x_M) = \frac{1}{\pi}\int_0^\pi \cos(n\tau - x_1\sin\tau - x_2\sin 2\tau - ... - x_M\sin M\tau)d\tau \quad (6)$$

A related class of functions are the modified Bessel functions of the first kind (Abramowitz and Stegun 1972)

$$I_\alpha(x) = i^{-\alpha}J_\alpha(ix) \quad (7)$$

which are the solutions to the modified Bessel differential equation

$$x^2\frac{d^2y}{dx^2} + x\frac{dy}{dx} - (x^2 + \alpha^2)y = 0 \quad (8)$$

The corresponding modified generalized Bessel functions are defined by (Dattoli *et al.* 1991)

$$I_n(x_1, x_2, ...x_M) = \frac{1}{\pi}\int_0^\pi \cos(n\tau)\exp(\sum_{s=1}^M x_s\cos s\tau)d\tau \quad (9)$$

The following are some important properties of modified generalized Bessel functions (Dattoli and Torre 1996, Dattoli *et al.* 1992):

$$I_n(x_1, x_2, ...x_M) = I_{-n}(x_1, x_2, ...x_M) \quad (10)$$

$$I_n(x_1, x_2, ...x_M) \to 0 \quad \text{as} \quad n \to \infty \quad (11)$$

$$\sum_{n=-\infty}^\infty e^{in\tau}I_n(x_1, x_2, ...x_M) = \exp\left[\sum_{s=1}^M x_s\cos(s\tau)\right] \quad (12)$$

$$I_n(x_1, x_2, ...x_M) = \sum_{l=-\infty}^\infty I_{n-Ml}(x_1, x_2, ...x_{M-1})I_l(x_M) \quad (13)$$

These will be used in the derivation of the natural connectivity of regular ring lattices in the next section.

## 3. Natural Connectivity of Regular Ring Lattices

A regular ring lattice $R_{N,K}$ is a $2K$-regular graph with $N$ vertices in a ring in which each vertex is connected to its $2K$ neighbors ($K$ on either side). They have also been called *K*-cycles and pristine worlds (Barahona and Pecora 2002) as they are the starting point for the small-world construction (Watts and Strogatz 1998). The adjacency matrix $A$ of $R_{N,K}$ is a circulant matrix in the form

$$A = \begin{bmatrix} c_0 & c_1 & ... & c_{N-1} \\ c_{N-1} & c_0 & ... & c_{N-2} \\ ... & ... & ... & ... \\ c_1 & c_2 & ... & c_0 \end{bmatrix} \quad (14)$$

where $c_k = 0$ if $\pm k \mod N > K$ or $k = 0$, otherwise $c_k = 1$. We show an example of a regular ring lattice in Fig. 1.

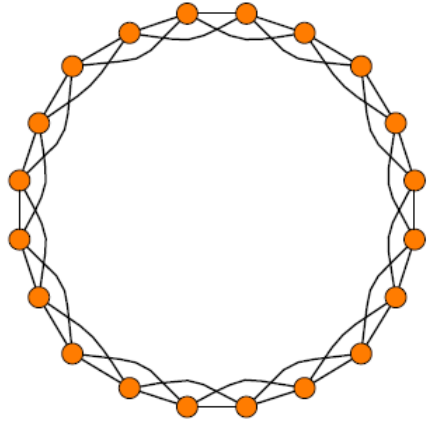

Fig. 1 Regular ring lattice $R_{N,K}$ with $N = 20$ and $K = 2$.

The eigenvectors of a circulant matrix are the columns of the discrete Fourier transform matrix of the same size. Consequently, the eigenvalues of a circulant matrix can be obtained by the Fast Fourier transform (Gray 2006, Davis 1979) as

$$\lambda_j = \sum_{k=0}^{N-1} c_k \exp\left(-\frac{2\pi i k(j-1)}{N}\right), \quad j = 1, 2...N \quad (15)$$

where $i = \sqrt{-1}$. For the regular ring lattices $R_{N,K}$, the spectrum is then given by

$$\begin{aligned} \lambda_j &= \sum_{k=1}^{K} \exp\left(-\frac{2\pi i k(j-1)}{N}\right) + \sum_{k=N-K}^{N-1} \exp\left(-\frac{2\pi i k(j-1)}{N}\right) \\ &= 2\sum_{k=1}^{K} \cos\left(\frac{2\pi k(j-1)}{N}\right) \end{aligned} \quad (16)$$

which can be substituted into Eq. (3) to obtain the expression of the natural connectivity

$$\bar{\lambda}_{R_{N,2K}} = \ln\left[\frac{1}{N}\sum_{j=1}^{N}\exp\left(\sum_{k=1}^{K} 2\cos\left(\frac{2\pi k(j-1)}{N}\right)\right)\right] \quad (17)$$

Using Eq.(12), we can rewrite Eq. (17) as

$$\bar{\lambda}_{R_{N,2K}} = \ln\left[\frac{1}{N}\sum_{j=1}^{N}\sum_{n=-\infty}^{\infty}\exp\left(in\frac{2\pi(j-1)}{N}\right)I_n(\overbrace{2,2,...2}^{K})\right]$$
$$= \ln\left[\frac{1}{N}\sum_{n=-\infty}^{\infty}I_n(\overbrace{2,2,...2}^{K})\sum_{j=1}^{N}\exp\left(in\frac{2\pi(j-1)}{N}\right)\right] \quad (18)$$

Note that

$$\frac{1}{N}\sum_{j=1}^{N}\exp\left(i\frac{2n\pi(j-1)}{N}\right) = \frac{1}{N}\sum_{j=1}^{N}\cos\left(\frac{2n\pi(j-1)}{N}\right) + i\frac{1}{N}\sum_{j=1}^{N}\sin\left(\frac{2n\pi(j-1)}{N}\right)$$
$$= \delta_{n,Nt} \quad (19)$$

where $\delta_{n,Nt}, t\in\mathbb{Z}$ is the Kronecker delta, i.e., $\delta_{n,Nt}=1$ if $n=Nt$, and $\delta_{n,Nt}=0$ otherwise. Then Eq. (18) simplifies to

$$\bar{\lambda}_{R_{N,K}} = \ln\left(\sum_{n=-\infty}^{\infty}I_n(\overbrace{2,2,...2}^{K})\cdot\delta_{n,Nt}\right) = \ln\left(2\sum_{t=1}^{\infty}I_{Nt}(\overbrace{2,2,...2}^{K}) + I_0(\overbrace{2,2,...2}^{K})\right) \quad (20)$$

The asymptotic convergence shown in Eq. (11) implies that $I_{Nt}(\overbrace{2,2,...2}^{K})\to 0$ as $N\to\infty$, which leads to the following asymptotic result.

**Theorem 1** Let $R_{N,K}$ be a regular ring lattice, then the natural connectivity of $R_{N,K}$ is

$$\bar{\lambda}_{R_{N,K}} = \ln\left(I_0(\overbrace{2,2,...2}^{K}) + o(1)\right) \quad (21)$$

where $o(1)\to 0$ as $N\to\infty$.

Using the properties shown in Eq. (13) and Eq. (10), we then obtain that

$$I_0(\overbrace{2,2,...2}^{K}) = \sum_{l=-\infty}^{\infty}I_{Kl}(\overbrace{2,2,...2}^{K-1})I_l(2) = I_0(\overbrace{2,2,...2}^{K-1})I_0(2) + 2\sum_{l=1}^{\infty}I_{Kl}(\overbrace{2,2,...2}^{K-1})I_l(2) \quad (22)$$

Note that $\sum_{l=1}^{\infty}I_{Kl}(\overbrace{2,2,...2}^{K-1})I_l(2) > 0$ and $I_0(2) = 2.2796 > 1$ whence it follows that

$$I_0(\overbrace{2,2,...2}^{K}) > I_0(\overbrace{2,2,...2}^{K-1}) \quad (23)$$

Thus $\bar{\lambda}_{R_{N,K}} > \bar{\lambda}_{R_{N,K-1}}$, i.e., the natural connectivity of regular ring lattices $R_{N,K}$ increases monotonically with $K$. The implication is that as we increase the number of neighbours in regular ring lattices the robustness of the graph increases.

Note that a cycle graph $C_N$, which consists of a single cycle, is a special regular ring lattice with $K=1$, i.e., $C_N = R_{N,1}$, we obtain the following corollary:

**Corollary 1** Let $C_N$ be a cycle graph, then the natural connectivity of $C_N$ is

$$\bar{\lambda}_{C_N} = \ln\left(I_0(2)+o(1)\right) \tag{24}$$

where $o(1) \to 0$ as $N \to \infty$.

Using the recurrence property shown in Eq. (13), we can devise a recursive analytical procedure to calculate $I_0(\overbrace{2,2,...2}^{K})$ in terms of standard Bessel functions $I_n(x)$. In Fig. 2, we show both the numerical and analytical results and find that the natural connectivity of regular ring lattices increases with $K$ monotonically, and that the natural connectivity is independent of the network size when $N$ is large. This agrees well with the analytical results.

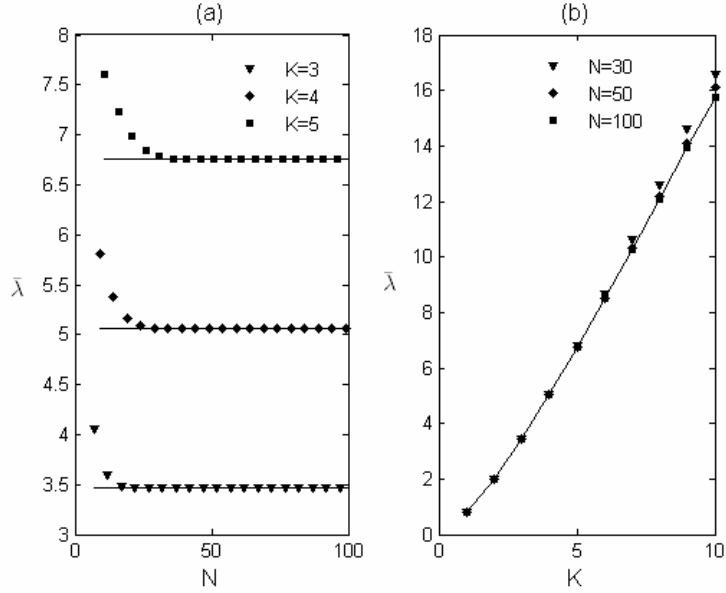

Fig. 2. Natural connectivity of regular ring lattices, $R_{N,K}$: (a) $\bar{\lambda}$ vs. $N$ with $K=3,4,5$ from bottom to top; (b) $\bar{\lambda}$ vs. $K$ with $N=30,50,100$. The symbols represent numerical results and lines represent the asymptotic analytical results as $N \to \infty$ according to Eq. (21).

For the sake of comparison, we show in Fig. 3 the algebraic connectivity $a$ of regular ring lattices. The algebraic connectivity is the first non zero eigenvalue of the Laplacian matrix of the graph and has been widely used as a measure of the connectedness of the network and the resistance to edge deletion. Our results show that the algebraic connectivity of regular ring lattices increases with $K$ monotonically, but decreases with $N$ monotonically. In fact, the algebraic connectivity approaches zero when $N$ is large. Our intuitive understanding of robustness suggests that the robustness of a regular ring lattice with given $K$ should not depend on the size of the graph $N$, when $N$ is large. Therefore the natural connectivity as a measure of robustness agrees with our intuition, while the algebraic connectivity does not.

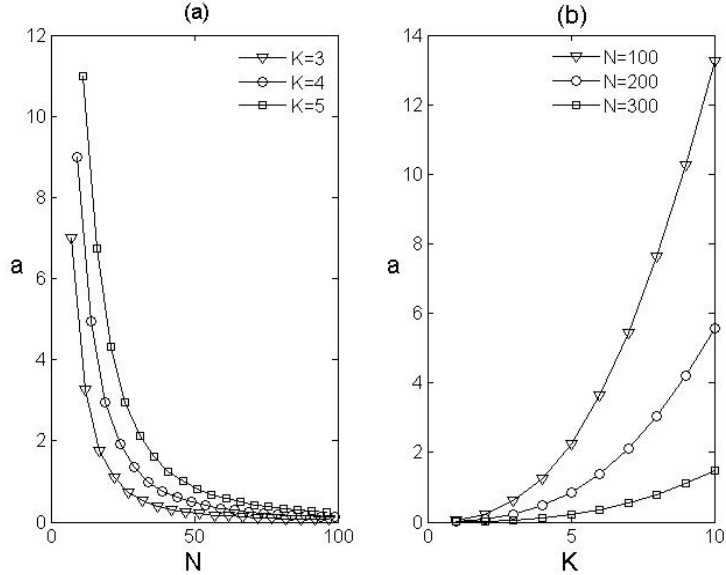

Fig. 3. Algebraic connectivity $a$ of regular ring lattices, $R_{N,K}$: (a) $a$ vs. $N$ with $K=3,4,5$ from bottom to top; (b) $a$ vs. $K$ with $N=100, 200, 300$.

## 4. Natural Connectivity of Random Regular Graphs

To explore the natural connectivity of regular ring lattices in depth, we randomize regular ring lattices and generate random regular graphs through a degree-preserving, random rewiring procedure. It is difficult to obtain the eigenvalues of random regular graphs analytically. In this paper, we present numerical results obtained from Eq. (3). The rewiring algorithm is as follows:

(1) Choose at random two edges denoted by $(v_1, w_1)$ and $(v_2, w_2)$;

(2) Replace the two edges with two new ones $(v_1, v_2)$ and $(w_2, w_2)$.

Note that the above process could lead to multi-edges or self-loops. We avoid that by rejecting any rewirings that lead to non-simple graphs.

In Fig. 4 we show the natural connectivity as a function of the number of rewirings starting from regular ring lattices with different $N$ and $K$. We find that the natural connectivity decreases with the process of random degree-preserving rewirings and then achieves a steady value. This means that random regular graphs are less robust than regular ring lattices reaching an asymptotic value for a totally randomized regular graph.

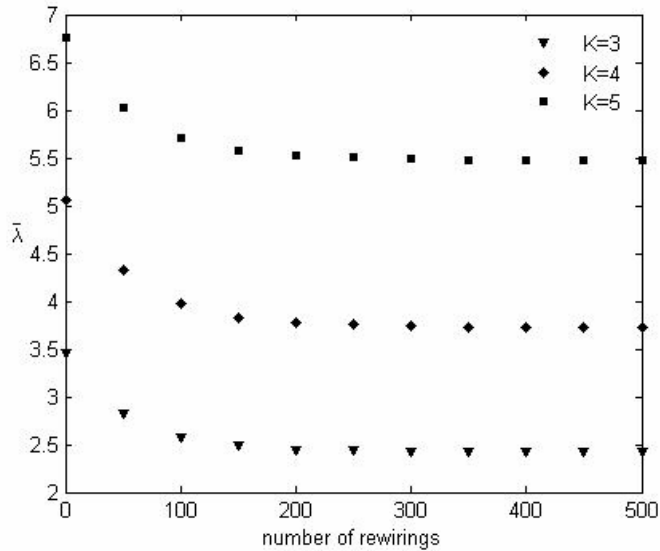

Fig. 4 Natural connectivity as a function of the number of random degree-preserving rewiring starting from regular ring lattices with $K = 3, 4, 5$ where $N = 100$. Each quantity is an averaged over 100 realizations.

We show in Fig. 5 the steady value of natural connectivity of random regular graphs as a function of $N$ and $K$ respectively. We find that the natural connectivity of random regular graphs decreases with $N$ and increases with $K$. For the sake of comparison, we also show the natural connectivity of regular ring lattices with the same number of vertices and edges. It is clear that regular ring lattices are more robust than random regular graphs.

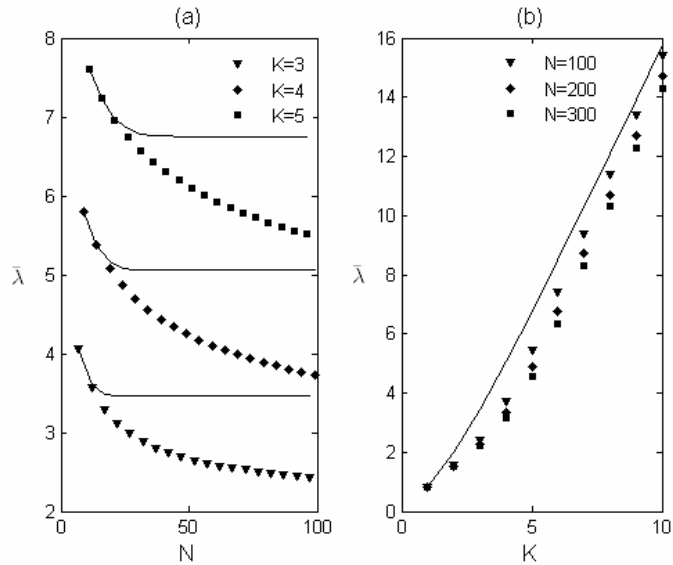

Fig. 5 Natural connectivity of random regular graphs: (a) $\bar{\lambda}$ vs. $N$ with $K = 3, 4, 5$. The solid lines represent the natural connectivity of regular ring lattices; (b) $\bar{\lambda}$ vs. $K$ with $N = 100, 200, 300$. Each quantity is an average over 100 realizations. The solid line here represents the asymptotic behavior as $N \to \infty$ of regular ring lattices.

## 5. Conclusions

We have investigated the natural connectivity of regular graphs in this paper. We have shown that the natural connectivity of regular ring lattices is asymptotically independent of network size and it increases monotonically with the degree. We have generated random regular graphs by degree-preserving, random rewirings from regular ring lattices and have demonstrated that the natural connectivity of random regular graphs is smaller than that of regular ring lattices. For these regular graphs, randomness decreases the robustness of the network.

## Acknowledgment

This work is in part supported by the National Science Foundation of China under Grant No. 60904065, No. 70501032 and No. 70771111.

Attack Information Parameter" *Chin Phys Lett,* 24, pp. 2138-2141, 2007a.

J. Wu, H. Z. Deng, Y. J. Tan & D. Z. Zhu, "Vulnerability of complex networks under intentional attack with incomplete information" *J Phys A,* 40, pp. 2665-2671, Mar 2007b.

J. Wu, Y.-J. Tan, H.-Z. Deng, Y. Li, B. Liu & X. Lv, "Spectral Measure of Robustness in Complex Networks" *arXiv: 0802.2564*.
18